\documentclass[a4paper]{jpconf}
\usepackage{graphicx}
\begin{document}
\title{On the statistical mechanics of shape fluctuations of nearly spherical lipid vesicle}

\author{I Bivas and N S Tonchev}

\address{Institute of Solid State Physics, Bulgarian Academy of Sciences, 72~Tzarigradsko chaussee blvd., Sofia~1784, Bulgaria}

\ead{bivas@issp.bas.bg, tonchev@issp.bas.bg}

\begin{abstract}
The mechanical properties of  biological membranes play an important role in the structure and the functioning of living organisms. One of the most widely used methods for determination of the bending elasticity modulus of the model lipid membranes (simplified models of the biomembranes with similar mechanical properties) is analysis of the shape fluctuations of the nearly spherical lipid vesicles. A theoretical basis of such an analysis is developed by Milner and Safran. In the present studies we analyze their results using an approach based on the Bogoljubov inequalities and the approximating Hamiltonian method. This approach is in accordance with the principles of statistical mechanics and is free of contradictions. Our considerations validate the results of Milner and Safran if the stretching elasticity $K_s$ of the membrane tends to zero.
\end{abstract}

\section{\label{s1}Introduction}
The biological membranes are one of the most important building blocks of living matter. Their mechanical properties determine to a great extent their structure and functioning. This is the reason for the interest in these properties over the years \cite{Sei97,Saf99,Nag12,Dim14}.

According to the model of Singer and Nicolson~\cite{Sin72}, the biomembrane consists of a lipid bilayer, in which integral proteins float. Evidently, in the frames of this model, the mechanical properties of the biomembrane are tightly connected with those of the lipid bilayer. In addition, the lipid bilayer is a well-defined structure that can be used in physical experiments. This is why the study of the mechanical properties of the lipid bilayers is continuously growing.

One of the most widely used methods for determination of the bending elasticity of a lipid bilayer is the analysis of the shape fluctuations of nearly spherical lipid vesicles~\cite{Mil87,Mel97}. The aim of the present paper is to reconsider the theoretical basis of this analysis, proposed by Milner and Safran~\cite{Mil87}.

\section{\label{s2}Mechanical energy stored in the shape fluctuations of a nearly spherical lipid vesicle}

Following Helfrich~\cite{Helf73}, we consider a small patch of the lipid bilayer with area $\Delta s$, tension $\sigma$, and area in its tension free state $\Delta s^{tf}$. Let $c_1$ and $c_2$ be the main curvatures of the patch under consideration. If the patch is tension-free then the bending energy density  $g_c^0$ can be
written in the following form:
\begin{eqnarray}
g_c^0(c_1,c_2)=\frac{1}{2}K_c(c_1+c_2-c_0)^2+\overline{K}_cc_1c_2, \label{h10}
\end{eqnarray}
where $K_c$ is the bending elasticity of the bilayer, $c_0$ is the spontaneous curvature of the membrane, and $\overline{K}_c$ is the saddle splay bending elasticity. In the present paper only symmetrical membranes (with $c_0=0$) will be considered. The bending energy $\Delta G_c$ of the whole patch is:
\begin{equation}
\Delta G_c = g_c^0(c_1,c_2)\Delta s^{tf}. \label{h13a}
\end{equation}

If the patch is not tension free then its stretching energy density $g_s$ is expressed via its tension $\sigma$ (here we assume that $\sigma$ is a constant all over the membrane \cite{Biv02}) as:
\begin{equation}
g_s=\frac{1}{2}\frac{\sigma^2}{K_s}, \label{h15}
\end{equation}
where $K_s$ is the stretching elasticity of the bilayer. The stretching energy $\Delta G_s$ of the patch is:
\begin{equation}
\Delta G_s =g_s\Delta s^{tf}. \label{h15a}
\end{equation}

Let $\Delta G$ be the total mechanical energy of of the patch:
\begin{equation}
\Delta G=\Delta G_c+\Delta G_s. \label{h16}
\end{equation}
The total deformation energy $G(t)$ of the vesicle is obtained by
integration of $\Delta G(t)$ on the vesicle surface $S(t)$:
\begin{equation}
G(t)=\oint_{S(t)}\Delta G(t). \label{h16a}
\end{equation}
In the last equation the contribution of the saddle-splay elasticity can be disregarded. This is due to the Gauss-Bonnet theorem which assures that if the topology of the vesicle does not change then the contribution of this elasticity does not depend on the shape fluctuations of the vesicle.

Let us consider a nearly spherical lipid vesicle. The volume $V$ of the vesicle is assumed not to be fluctuating. Let $R_0$ be the radius of a sphere with a volume of $V$. Let the origin $O$ of a laboratory reference frame be placed inside the vesicle. A point on the surface of the vesicle with polar coordinates $(\theta,\varphi)$ is chosen. Let $R(\theta,\varphi)$ be the modulus of the radius-vector at this point. The dimensionless quantity $u(\theta,\varphi,t)$ is defined by the equation:
\begin{equation}
R(\theta,\varphi,t)=R_0[1+u(\theta,\varphi,t)], \label{e5}
\end{equation}
where $t$ is the time variable. The function $u(\theta, \varphi,t)$ is decomposed in a series as follows:
\begin{equation}
u(\theta,\varphi,t) =
\sum_{n=0}^{n_{max}}\sum_{m=-n}^{n}
u_n^m(t) Y_n^m(\theta,\varphi), \label{e6}
\end{equation}
where $Y_n^m(\theta,\varphi)$ is the orthonormal basis (for simplicity chosen real) of the spherical harmonics functions~\cite{Jan60}. A cut-off $n_{max}\sim R_0/\lambda$ is introduced in the sum, where $\lambda$ is of the order of the intermolecular distance. As the harmonics with indexes $n=1$ and $m=-1,0,1$ correspond to pure translation of the vesicle, the origin $O$ can be chosen in a way that $u_1^m=0$. Because of the requirement for volume conservation the amplitude $u_0^0(t)$ can be expressed
as~\cite{Mil87}:
\begin{equation}
u_0^0(t)=-\frac{1}{2\pi^{1/2}}\sum_{n=2}^{n_{max}}\sum_{m=-n}^{n}
[u_n^m(t)]^2. \label{e7}
\end{equation}

\section{\label{s3}Theory of Milner and Safran}
According to the theory, developed by Milner and Safran \cite{Mil87}, the mechanical energy $G(t)$ of the vesicle is expressed as follows:
\begin{equation}
G(t)= \sum_{n=2}^{n_{max}}\sum_{m=-n}^{n}\frac{1}{2}K_c (n-1)(n+2)[n(n+1)+ \overline{\sigma}][u_n^m(t)]^2 \label{h19}
\end{equation}
where:
\begin{equation}
\overline{\sigma} = \frac{\sigma (R_0)^2}{K_c} \label{h20}
\end{equation}
is a dimensionless parameter.
The quantity $\sigma$ (with dimension of tension) is considered by Milner and Safran as a Lagrange multiplier, not fluctuating with time, which ensures the mean area of the vesicle membrane to be equal to some prescribed value.
The energy $G(t)$ from Eq.~\eref{h19} is a sum of the energies of not interacting oscillators. This energy does not depend on the stretching elasticity $K_s$ of the membrane. It is a function  only of $K_c$, $R_0$, and $\sigma$.

In the frames of this theory, the amplitudes $u_n^m$ of the different fluctuation modes are not time correlated. For $n\ge 2$ the time mean squares $ \overline{ [u_n^m(t)]^2}$ are:
\begin{equation}
\overline{ [u_n^m(t)]^2} = \frac{kT}{K_c}
\frac{1}{(n-1)(n+2)[n(n+1)+\overline{\sigma}]} \label{h21}
\end{equation}
where $kT$ is the Boltzmann factor.

\section{\label{s4}The Model Hamiltonian}

Let us denote with $H(U)\equiv H(u_2^{-2}, u_2^{-1},\dots, u_{n_{max}}^{n_{max}})$ the model Hamiltonian  describing a fluctuating vesicle. The symbol  $U$ is used as shorthand for the real value  functions $(u_2^{-2}, u_2^{-1},\dots, u_{n_{max}}^{n_{max}})$ that are the spherical harmonic amplitudes (see~\eref{e6}),
appearing in the the expansion of the vesicle shape fluctuations from the equivalent volume sphere with radius $R_0$. From here on the notation $u^l_m$ is used as an abbreviation of $u^l_m$(t). We assume that our system is described by a Hamiltonian $H(U)$, consisting in a sum of the  bending $H_c(U)$ and stretching  $H_s(U)$ energies:
\begin{equation}
H(U)=H_c(U)+H_s(U),
\label{IB}
\end{equation}
where
\begin{equation}
H_c(U)=\frac{1}{2}K_c \sum_{n=2}^{n_{max}}\sum_{m=-n}^n (n-1)n(n+1)(n+2)(u_n^m)^2
\end{equation}
and
\begin{equation}
H_s(U)=\frac{1}{2}\frac{[\sigma(U)]^2}{K_s}S_0^0.
\label{Hs1}
\end{equation}
In the last equation $S_0^0$ is the area of of the vesicle membrane when it is tension free.  The membrane tension $\sigma=\sigma(U)$, which is assumed  constant all over the membrane, is given by the expression:
\begin{equation}
\sigma(U) = K_s\frac{4\pi(R_0)^2+\Delta S(U)-S_0^0}{S_0^0}.
\label{r17}
\end{equation}
The quantity $\Delta S(U)$ is the excess area of the vesicle (the difference between the area of the vesicle membrane and the area of the sphere with a volume equal to
that of the vesicle). It is expressed in terms of the amplitudes $u_n^m$ in the following way~\cite{Mil87}:
\begin{equation}
\Delta S(U)=  \frac{(R_0)^2}{2}
\Bigg[ \sum_{n=2}^{n_{max}}\sum_{m=-n}^n
(n-1)(n+2) (u_n^m)^2 \Bigg]. \label{e18}
\end{equation}
For further application it is convenient to present the model Hamiltonian $H(U)$ in the following  equivalent form:
\begin{eqnarray}
H(U) &=& \frac{1}{2}K_c \sum_{n=2}^{n_{max}}\sum_{m=-n}^n \Bigg(
(n-1)(n+2) \nonumber \\
&\times& \Bigg\{n(n+1)+\sigma_0+\frac{K_s}{2K_c}
\frac{(R_0)^2}{S_0^0}\Bigg[\Delta S(U) \Bigg]\Bigg\}(u_n^m)^2\Bigg), \label{e120}
\end{eqnarray}
where:
\begin{equation}
\sigma_0=\frac{(R_0)^2}{K_c}K_s\,\frac{4\pi(R_0)^2-S_0^0}{S_0^0}. \label{e16}
\end{equation}
Eq.~\eref{e120} gives us a hint to introduce the key quantity:
\begin{equation}
\hat{\sigma}= \sigma_0+ \frac{K_s}{2K_c}
\frac{(R_0)^2}{S_0^0}\langle \Delta S(U) \rangle_{H(U)},  \label{e21}
\end{equation}
where the symbol $\langle ... \rangle_{H(U)}$ denotes a thermodynamic average calculated with the model Hamiltonian $H(U)$:
\begin{equation}
\langle \dots \rangle_{H(U)}=\left\{Z[H(U)]\right\}^{-1}\int dU \dots \exp{\left[-\frac{H(U)}{kT}\right]},
\end{equation}
where
\begin{equation}
Z[H(U)] =\int dU \exp{\left[-\frac{H(U)}{kT}\right]} 
\end{equation}
is the statistical sum of the model.
Using Eq.~\eref{e21} we can rewrite Eq.~\eref{e120} in the form:
\begin{eqnarray}
H(U) & =&  \frac{1}{2}K_c  \sum_{n=2}^{n_{max}}\sum_{m=-n}^n
\Bigg\{ (n-1)(n+2) \nonumber \\
& \times& \Bigg[ n(n+1)+\hat{\sigma}+\frac{K_s}{2K_c}
\frac{(R_0)^2}{S_0^0}\bigg(\Delta S(U)- \langle \Delta S(U) \rangle_{H(U)}\bigg) \Bigg](u_n^m)^2\Bigg\}. \label{e22}
\end{eqnarray}
Let us note that if we drop the third term in the rectangular  bracket in the rhs of Eq.~\eref{e22} and replace $\overline{\sigma}$ from Eq.~$\eref{h20}$  with  $\hat{\sigma}$, we obtain a result coinciding in form with the one of Milner and Safran:
\begin{eqnarray}
\langle (u_n^m)^2\rangle_{H_{MS}(U,\hat{\sigma})} 
= \frac{kT}{K_c} \frac{1}{(n-1)(n+2)[n(n+1)+\hat{\sigma}]}. \label{e67ab}
\end{eqnarray}
The relevant Hamiltonian in the calculation of the thermodynamic mean value in the lhs of Eq.~\eref {e67ab} turns out to be the "Milner and Safran Hamiltonian" $H_{MS}(U,\hat{\sigma})$ of a system, consisting of independent harmonic oscillators, of the kind:
\begin{eqnarray}
H_{MS}(U,\hat{\sigma}) =  \frac{1}{2}K_c  \sum_{n=2}^{n_{max}}\sum_{m=-n}^n
\Bigg\{ (n-1)(n+2)
 \Bigg[ n(n+1)+\hat{\sigma} \Bigg](u_n^m)^2\Bigg\}. \label{e22a}
\end{eqnarray}
A sufficient condition this step to be acceptable is the smallness (in some sense) of  the term
\begin{equation}
\frac{K_s}{2K_c}
\frac{(R_0)^2}{S_0^0}\bigg[\Delta S(U)- \langle \Delta S(U) \rangle_{H(U)}\bigg]
\label{sc}.
\end{equation}
Without entering in details we note that the smallness of the above expression is equivalent to the condition for smallness of the excess area fluctuations.

To avoid confusion, we warn the reader that Eq.~\eref{e67ab} can be obtained from Eq.~\eref{h21} by the replacement of $\overline{\sigma}$ with
$\hat{\sigma}$. The quantity $\overline{\sigma}$ has been introduced in the theory by hand to ensure that the mean excess
area of the fluctuating membrane is equal to some prescribed value, while the quantity $\hat{\sigma}$ has a self-consistent origin depending on the model Hamiltonian itself.

It can be concluded that the approach of Milner and Safran employs  the implicit assumption that the term~\eref{sc} can be neglected in the model Hamiltonian $H(U)$. At this stage let us point out that an exact treatment of the thermodynamics of the model Hamiltonian $H(U)$ seems to be unrealistic due to
the interaction between the modes in the the stretching energy term. Indeed, it is very hard to obtain additional results beyond the Milner and Safran ones concerning the case when modes do not interact, unless some approximation tricks or variational methods are used.

\section{\label{s5}The approximating Hamiltonian}
In this section we show that the approximating Hamiltonian  $H_{app}(U,\Sigma)$ of the kind:
\begin{equation}
H_{app}(U,\Sigma)  =  \frac{1}{2}K_c  \sum_{n=2}^{n_{max}}\sum_{m=-n}^n
\{ (n-1)(n+2) [ n(n+1)+\Sigma ](u_n^m)^2\},
\label{e122}
\end{equation}
where $\Sigma$ is an appropriately defined trial quantity, is the best approximation of the model Hamiltonian $H(U)$ by a Hamiltonian of a system, consisting in independent harmonic oscillators. Hereafter the problem is to obtain $\Sigma$. To this end we use the Bogoljubov variational principle~\cite{Bra00}. In order to use it, some preliminary mathematical manipulations on the model Hamiltonian $H(U)$ need to be performed.

First, we introduce the functions ${\cal A}(U)$ and ${\cal T}(U,\sigma_0)$ in the following way:
\begin{equation}
{\cal A}(U) =\left(\frac{K_s}{2S_0^0}\right)^{1/2}\frac{(R_0)^2}{2}  \sum_{n=2}^{n_{max}}\sum_{m=-n}^n(n-1)(n+2)(u_n^m)^2 \label{e39}
\end{equation}
and
\begin{equation}
{\cal T}(U,\sigma_0)=\frac{1}{2} K_c \sum_{n=2}^{n_{max}}\sum_{m=-n}^n(n-1)(n+2)[n(n+1)+\sigma_0](u_n^m)^2. \label{e140}
\end{equation}

Then we define a trial Hamiltonian $H_{app}(U,X,\sigma_0)$ as:
\begin{equation}
H_{app}(U,X,\sigma_0)= {\cal T}(U,\sigma_0) +2X{\cal A}(U)-X^2, \label{e44}
\end{equation}
where $X$ is an arbitrary real number. The Hamiltonian $H_{app}(U,X,\sigma_0)$, obtained in this way, is linearized with respect to the
squares  of the amplitudes $(u_n^m)^2$. It is easy to see that:
\begin{equation}
0\leq [{\cal A}(U)-X]^2=H(U)-H_{app}(U,X,\sigma_0)  \label{e45}
\end{equation}
for arbitrary $X$.
From Eqs.~\eref{e39}, \eref{e140}, and \eref{e44} we obtain
\begin{eqnarray}
H_{app}(U,X,\sigma_0) &=& \frac{1}{2}K_c  \sum_{n=2}^{n_{max}}\sum_{m=-n}^n
\Bigg\{ (n-1)(n+2) \nonumber \\
&\times &\Bigg[ n(n+1)+\sigma_0
+\left(\frac{2K_s}{S_0^0}\right)^{1/2}\frac{(R_0)^2}{K_c}X\Bigg]\Bigg\}
(u_n^m)^2 - (X)^2. \label{e145}
\end{eqnarray}
According to the Bogolyubov inequalities (see e.~g.~\cite{Bra00}), for all $X$ it is true that:
\begin{eqnarray}
\langle H(U)-H_{app}(U,X,\sigma_0) \rangle_{H(U)} & \le & f[H(U)]- f[H_{app}(U,X,\sigma_0)] \nonumber \\
&\le & \langle H(U)-H_{app}(U,X,\sigma_0) \rangle_{H_{app}(U,X,\sigma_0)} \label{e46}
\end{eqnarray}
where $f[H(U)]=-kT\ln Z[H(U)]$ is the free energy of the system with Hamiltonian $H(U)$, $\langle H(U)-H_{app}(U,X,\sigma_0) \rangle_{H(U)}$ is the thermodynamic average calculated with the Hamiltonian $H(U)$, and
$f[H_{app}(U,X,\sigma_0)]=-kT\ln Z[H_{app}(U,X,\sigma_0)]$ and $\langle H(U)-H_{app}(U,X,\sigma_0) \rangle_{H_{app}(U,X,\sigma_0)}$ are a free energy and a thermodynamic average, calculated with $H_{app}(U,X,\sigma_0)$.
Since the residual Hamiltonian $H(U)-H_{app}(U,X,\sigma_0)$ is non-negative (see Eq.~\eref{e45}), the application of the Bogolyubov inequalities~\eref{e46} yields that for all $X$:
\begin{equation}
0 \le f[H(U)] - f[H_{app}(U,X,\sigma_0)] \le \langle [{\cal A}(U)-X]^2 \rangle_{H_{app}(U,X,\sigma_0)}. \label{e148}
\end{equation}
The best approximation of the free energy of the model system with Hamiltonian $H_{app}(U,X,\sigma_0)$ (see Eq.~\eref{e145}) is given by the absolute maximum principle:
\begin{equation}
\max_X f[H_{app}(U,X,\sigma_0)]=f[H_{app}(U,\overline{X},\sigma_0)]. \label{e149}
\end{equation}
We note that $f[H_{app}(U,X,\sigma_0)]$  attains its maximum with respect to  the parameter $X$ at the solution of the equation:
\begin{equation}
\frac{\partial f[H_{app}(U,X,\sigma_0)]}{\partial X}=0,
\end{equation}
which yields
\begin{equation}
\langle {\cal A}(U) \rangle_{H_{app}(U,X,\sigma_0)}=X.
\label{1sce}
\end{equation}
This is a self-consistency  equation for the parameter $X$. It can be shown that this (self-consistency) equation has only one solution, namely $\overline{X}$. In its explicit form this equation takes the form:
\begin{equation}
X=\frac{kT\sigma_1}{4}\sum_{n=2}^{n_{max}} \frac{2n+1}{n(n+1)+ \sigma_0+\sigma_1X}. \label{e68a}
\end{equation}

The absolute maximum condition ~\eref{e149} and inequalities \eref{e46} impose  the
following bounds on the free energy difference $f[H(U)] -
f[H_{app}(U,X,\sigma_0)]:$
\begin{equation}
0 \le f[H(U)]- f[H_{app}(U,\overline{X},\sigma_0)] \le \langle [{\cal A}(U)-\overline{X}]^2
\rangle_{H_{app}(U,\overline{X},\sigma_0)}. \label{e50}
\end{equation}

Evidently $H_{app}(U,\overline{X},\sigma_0)$
provides the best  approximation from below of the free energy of the model Hamiltonian $H(U)$ (see Eq.~\eref{e22}) with the free energy of the approximating Hamiltonian
\begin{equation}
H_{app}(U,\overline{X},\sigma_0)= {\cal T}(U,\sigma_0) +2\overline{X}{\cal A}(U)-\left(\overline{X}\right)^2.
\label {AH1}
\end{equation}



Using Eqs.~\eref{e39} and \eref{e140}, the approximating Hamiltonian $H_{app}(U,\overline{X},\sigma_0)$, describing an ensemble of
noninteracting oscillators, can be rewritten in the more convenient form:
\begin{equation}
H_{app}(U,\overline{X},\sigma_0)=\frac{1}{2}K_c  \sum_{n=2}^{n_{max}}\sum_{m=-n}^n
(n-1)(n+2) \left[ n(n+1)+\tilde{\sigma}\left(\sigma_0,\overline{X}\right)\right](u_n^m)^2 - \left(\overline{X}\right)^2, \label{e57}
\end{equation}
where
\begin{equation}
\tilde{\sigma}\left(\sigma_0,X \right)= \sigma_0+\left(\frac{2K_s}{S_0^0}\right)^{1/2}
\frac{(R_0)^2}{K_c}X \equiv \sigma_0 + \sigma_1 X \label{e58}
\end{equation}
and
\begin{equation}
\sigma_1= \left(\frac{2K_s}{S_0^0}\right)^{1/2}\frac{(R_0)^2}{K_c}. \label{e59}
\end{equation}

The comparison of Eq.~\eref{e57} with Eq.~\eref{e122} shows that  $\tilde{\sigma}\left(\sigma_0,\overline{X} \right)$ plays the role of $\Sigma$ .


 To make the respective assessments it is necessary to have the analytical expressions for the free energy term (this is not difficult because the calculations are for  noninteracting harmonic oscillators) and to solve numerically the self-consistent equation in order to determine $\overline{X}$,
which depends on the quantities $kT$, $K_s$, $R_0$, and $K_c$, considered as fixed values for a definite vesicle.
The only free parameter in the theory remains $\sigma_{0}$.

Following our approach, the mean square value of the amplitude in Eq.~\eref{e67ab} is replaced by:
\begin{eqnarray}
\langle (u_n^m)^2\rangle_{H_{app}(U,\overline{X},\sigma_0)}  = \frac{kT}{K_c} \frac{1}{(n-1)(n+2)[n(n+1)+\tilde{\sigma}(\sigma_0,\overline{X})]}. \label{e67a}
\end{eqnarray}

In order to obtain Eq.~\eref{e67a} from Eq.\eref{e67ab} the following replacements need to be done:
\begin{equation}
\langle (u_n^m)^2\rangle_{H_{MS}(U,\hat \sigma)} \rightarrow \langle
(u_n^m)^2\rangle_{H_{app}(U,\overline{X},\sigma_0)},\qquad
\hat{\sigma}\rightarrow \tilde{\sigma}(\sigma_0,\overline{X}).
\end{equation}

Consequently,  in our approach the term 
\begin{equation}
H_s(U)=\frac{1}{2}\frac{[\sigma(U)]^2}{K_s}S_0^0
\label{Hs1a}
\end{equation}
ensures  that the mean square amplitudes $\langle
(u_n^m)^2\rangle_{H_{app}(U,\overline{X},\sigma_0)}$  are calculated by an
effective tension, appearing as solution of the following equation, deduced from Eq.~\eref{e68a}:
\begin{equation}
\Sigma = \sigma_0+\frac{kT\sigma_1^2}{4}\sum_{n=2}^{n_{max}}
\frac{2n+1}{n(n+1)+ \Sigma}.
\label{e68}
\end{equation}
The  solution of this equation is 
$\overline{\Sigma}=\tilde{\sigma}(\sigma_0,\overline{X})$. Clearly, $\overline{X}$ is function of $\sigma_0$, which means that $\tilde{\sigma}(\sigma_0,\overline{X})$, appearing as solution of Eq.~\eref{e68}, depends only on $\sigma_0$. In the Milner and Safran approach the corresponding quantity $\overline{\sigma}$ is an external parameter.

In order to  discuss the closeness of the free energies of the model and approximating systems, one must calculate the correlator $\langle [{\cal A}(U)-\overline{X}]^2
\rangle_{H_{app}(U,\overline{X},\sigma_0)}$ from Eq.~\eref{e148}. Since the Hamiltonian $H_{app}(U,\overline{X},\sigma_0)$ is linear with respect to the squares of the amplitudes $(u_n^m)^2$, after some lengthly but standard calculations we obtain:
\begin{equation}
\langle [{\cal A}(U)-\overline{X}]^2 \rangle_{H_{app}(U,\overline{X},\sigma_0)}=\frac{K_s}{S_0^0} \frac{(R_0)^4}{4} \Bigg[\frac{kT}{K_c} \Bigg]^2 \sum_{n=2}^{n_{max}} \frac{2n+1}{[n(n+1)+\tilde{\sigma}(\sigma_0,\overline{X})]^2}. \label{e661}
\end{equation}

Our further aim is to estimate the rhs of \eref{e661}. Its smallness provides a sufficient condition for the replacement of the Hamiltonian $H(U)$ with $H_{app}((U,\overline{X},\sigma_0))$.

\section{\label{s6}Numerical estimations and discussion}
If $\langle [{\cal A}(U)-\overline{X}]^2 \rangle_{H_{app}(U,\overline{X},\sigma_0)}$ is close enough to zero (see Eq.~\eref{e50}), we can replace the model Hamiltonian $H(U)$ with the approximating one $H_{app}(U,\overline{X},\sigma_0)$ and the model free energy $f[H(U)]$ with the approximating one $f[H_{app}(U,\overline{X},\sigma_0)]$.
To make numerical estimations, we use values of the quantities, participating in the developed by us theory that are typical for experiments where analysis of the shape fluctuations of nearly spherical lipid vesicles are carried out~\cite{Mel97}, namely: $K_s\sim100$~erg/cm$^2$; $K_c\sim10^{-12}$erg; $R_0\sim10^{-3}$cm; $S_0^0\sim 4\pi (R_0)^2\sim 1.256\times 10^{-5}$cm$^2$; $\sigma_1=4\times 10^9$erg$^{-0.5}$. With these values the correlator $\langle [{\cal A}(U)-\overline{X}]^2 \rangle_{H_{app}(U,\overline{X},\sigma_0)}$ from Eq.~\eref{e661}, normalized by the Boltzmann factor $kT$, can be presented in the form:
\begin{equation}
\frac{\langle [{\cal A}(U)-\overline{X}]^2 \rangle_{H_{app}(U,\overline{X},\sigma_0)}}{kT}=2\times 10^{4} \sum_{n=2}^{n_{max}} \frac{2n+1}{[n(n+1)+\tilde{\sigma}(\sigma_0,\overline{X})]^2}. \label{167}
\end{equation}
The use of the approximating Hamiltonian is justified when the correlator is small enough. We assume that this is true when the value of the correlator is less than the Boltzmann factor $kT$. With these numerical values this is fulfilled when $\overline{\Sigma} \ge 2.10^4$.

If we consider lower values of the stretching elasticity $K_s$, the lowest value
of $\overline{\Sigma}$, for which the use of the approximating Hamiltonian is  justifiable, decreases. In the limiting case $K_s \rightarrow 0$ (if
$\overline{\Sigma}$ is fixed), the approximating Hamiltonian can be used for all values of $\overline{\Sigma} $ satisfying the condition $\overline{\Sigma}  > -6$.
The last inequality is necessary in order to assure that the factors multiplying
the squares $(u_n^m)^2$ of the amplitudes $u_n^m$ of the fluctuation modes (see
Eq.~\eref{e57}) are finite and nonnegative.

\section{\label{s8}Conclusion}
In the present paper we show that results having the same functional dependences as those of Milner and Safran can be deduced, in accordance with the principles of statistical mechanics, by an approach based on the Bogolyubov inequalities and the approximating Hamiltonian method. 

It is proved that there is a value of the dimensionless factor $\overline{\Sigma} $, related to the tension of the membrane (see Eqs.~\eref{e58} and \eref{e68}), above which the application of our approach gives results that are  precise enough. This value is much greater than the values appearing in the experiments. The applicability of the results for the interval of typical experimental values of this quantity remains an open question.

From Eqs.~\eref{e148} and \eref{e661} it is clear that  when the stretching elasticity $K_s$ of the vesicle membrane tends to zero, keeping all the other quantities fixed, our theory becomes asymptotically exact.

\end{document}